\documentclass{PoS}

\usepackage{amsfonts,amssymb,amsmath}
\usepackage{graphicx}

\newcommand{\be}{\begin{equation}}
\newcommand{\bea}{\begin{eqnarray}}
\newcommand{\ee}{\end{equation}}
\newcommand{\eea}{\end{eqnarray}}
\newcommand{\bpi}{\begin{picture}}
\newcommand{\bce}{\begin{center}}
\newcommand{\epi}{\end{picture}}
\newcommand{\ece}{\end{center}}
\newcommand{\D}{\displaystyle}

\def\gb{{\rm I}\hspace{-0.07cm}\Gamma}

\def\g{\widetilde\gb}

\def\gb{{\rm I}\hspace{-0.07cm}\Gamma}
\def\g{\widetilde{{\rm I}\hspace{-0.07cm}\Gamma}}

\newcommand{\Valencia}{Departamento de F\'\i sica Te\'orica
and IFIC, Centro Mixto, Universidad de Valencia--CSIC \\
E-46100, Burjassot, Valencia, Spain}

\title{Infrared finite effective charge of QCD}

\ShortTitle{Infrared finite effective charge of QCD}

\author{A. C. Aguilar,$^a$ D.~Binosi$^{b}$
and  \speaker{J. Papavassiliou}$ ^a$\\
\llap{$^a$}\Valencia \\
\llap{$^b$}  European Centre for Theoretical Studies in Nuclear
  Physics and Related Areas (ECT*), \\Villa Tambosi, Strada delle
  Tabarelle 286, I-38050 Villazzano (TN), Italy. \\
E-mail: \email{Arlene.Aguilar@uv.es},\email{binosi@ect.it}, \email{Joannis.Papavassiliou@uv.es}}

\abstract{

We  show that  the gauge  invariant treatment  of  the Schwinger-Dyson
equations  of  QCD  leads  to  an infrared  finite  gluon  propagator,
signaling the  dynamical generation of an effective gluon mass, and
 a  non-enhanced ghost  propagator,  in  qualitative agreement  with
recent lattice data.   The truncation scheme employed is  based on the
synergy  between   the  pinch  technique  and   the  background  field
method. One of  its most powerful features is  that the transversality
of the gluon self-energy  is manifestly preserved, exactly as dictated
by the  BRST symmetry of the  theory.  We then explain,  for the first
time  in  the  literature,   how  to  construct  non-perturbatively  a
renormalization group invariant quantity out of the conventional gluon
propagator.   This  newly constructed  quantity  serves  as  the  natural
starting point  for defining  a non-perturbative effective  charge for
QCD,  which constitutes,  in  all respects,  the  generalization in  a
non-Abelian  context  of the  universal  QED  effective charge.   This
strong effective charge displays asymptotic freedom in the ultraviolet,
while in the low-energy regime  it freezes at a finite value, giving
rise  to an  infrared  fixed  point for  QCD.  Some possible  pitfalls
related to the extraction of  such an effective charge from infrared
finite gluon propagators, such as those found on the lattice, 
are briefly discussed.}

\FullConference{LIGHT CONE 2008 Relativistic Nuclear and Particle Physics\\
		 July 7-11  2008\\
		 Mulhouse, France}
\begin{document}

It  has been  known for  a long  time that  even though  the  gluon is
massless  at the  level  of the  fundamental  Lagrangian, and  remains
massless to all order in perturbation theory, the non-perturbative QCD
dynamics  generate  an  effective,  momentum-dependent  mass,  without
affecting    the   local    $SU(3)_c$   invariance,    which   remains
intact~\cite{Cornwall:1982zr}.    The  existence   of  this   mass  is
discovered by studying the Schwinger-Dyson equations (SDEs) of QCD, in
a  gauge-invariant framework known  as the  pinch technique  
(PT)~\cite{Cornwall:1982zr,Cornwall:1989gv}.
To obtain  massive
solutions it  is necessary to  include longitudinally-coupled massless
scalars  in the  Green's  functions,  which play  a  role rather  like
Goldstone  excitations, but  do not  signal any  sort of  breakdown of
local gauge  symmetry, which  is preserved.   Like standard
Goldstone bosons  these massless scalars  do not appear  explicitly in
the  $S$-matrix; however,  they play  a crucial  role  in confinement,
furnishing the required long range potential.

An effective low-energy field 
theory for describing the gluon mass   
is  the  gauged non-linear sigma model  known  as ``massive
gauge-invariant Yang-Mills''~\cite{Cornwall:1979hz}, with    
Lagrangian density  
\begin{equation}
{\cal L}_{MYM}= \frac{1}{2} G_{\mu\nu}^2 - 
m^2 {\rm Tr} \left[A_{\mu} - {g}^{-1} U(\theta)\partial_{\mu} U^{-1}(\theta) \right]^2\,,
\label{nlsm}
\end{equation}
where 
$A_{\mu}= \frac{1}{2i}\sum_{a} \lambda_a A^{a}_{\mu}$, the $\lambda_a$ are the SU(3) generators
(with  ${\rm Tr} \lambda_a  \lambda_b=2\delta_{ab}$), 
and the $N\times N$
unitary matrix $U(\theta) = \exp\left[i\frac{1}{2}\lambda_a\theta^{a}\right]$ 
describes the scalar fields $\theta_a$.  
Note that ${\cal L}_{MYM}$ is locally gauge-invariant under the combined gauge transformation 
\be
A^{\prime}_{\mu} = V A_{\mu} V^{-1} - {g}^{-1} \left[\partial_{\mu}V \right]V^{-1}\,, 
\qquad
U^{\,\prime} = U(\theta^{\,\prime}) = V U(\theta)\,,
\label{gtransfb}
\ee
for any group matrix $V= \exp\left[i\frac{1}{2}\lambda_a\omega^{a}(x)\right]$, where 
$\omega^{a}(x)$ are the group parameters. 
One might think that, by employing (\ref{gtransfb}), the fields  $\theta_a$ can always 
be transformed to zero, but this is not so if the $\theta_a$ contain vortices.
To use the ${\cal L}_{MYM}$ in (\ref{nlsm}), one solves the equations of motion for $U$ 
in terms of the gauge potentials and substitutes the result in the equations for the gauge potential.  
One then finds the Goldstone-like massless modes mentioned above.
This model admits  vortex
solutions~\cite{Cornwall:1979hz},  with a  long-range pure  gauge term  in  their potentials,
which endows  them with a topological quantum  number corresponding to
the center  of the gauge group  [$Z_N$ for $SU(N)$], and  is, in turn,
responsible for quark  confinement and gluon screening. 
Specifically, center vortices of  thickness $\sim m^{-1}$,  where $m$ is
the induced mass of the gluon, form a condensate because their entropy
(per  unit  size) is  larger  than  their  action.  This  condensation
furnishes an  area law to  the fundamental representation  
Wilson loop, thus confining quarks~\cite{Cornwall:1982zr,Cornwall:1979hz}. 

Of course ${\cal L}_{MYM}$ is not renormalizable, and breaks down in the ultraviolet.  
This breakdown simply reflects the fact that the gluon mass $m$ in (\ref{nlsm})
is assumed to be constant, while the solutions of the SDEs  
reveal that the mass is momentum-dependent, vanishing at large $q^2$~\cite{Cornwall:1982zr}.  
Specifically, when studying the SDE for the gluon propagator, $\Delta(q^2)$,  
one looks for infrared finite solutions, i.e. with  $\Delta^{-1}(0) > 0$. 
Such solutions may  
be  fitted  by     ``massive''  propagators  of   the form \mbox{$\Delta^{-1}(q^2)  =  q^2  +  m^2(q^2)$}, where  
$m^2(q^2)$ is  not ``hard'', but depends non-trivially  on the momentum  transfer $q^2$.
When the  renormalization-group 
logarithms are  properly taken into  account in the SDE analysis, one obtains,  in addition, 
the  non-perturbative  generalization  of  $\alpha(q^2)$, 
the  QCD  running  coupling (effective charge).
The presence of $m^2(q^2)$ in the argument of $\alpha(q^2)$
tames  the   Landau  singularity   associated   with  the
perturbative $\beta$  function, and the resulting  effective charge is
asymptotically free in  the ultraviolet , ``freezing'' at a  finite value in the infrared.

The general picture described above has received spectacular confirmation from 
a plethora of lattice studies, spanning a period of several years:
the gluon propagator  reaches indeed a finite (non-vanishing) value in the deep 
infrared, as predicted by Cornwall. This  rather characteristic  behavior 
was already seen in early studies~\cite{Alexandrou:2000ja}, and 
has been  firmly established  recently using
large-volume lattices, for both $SU(2)$~\cite{Cucchieri:2007md} and $SU(3)$~\cite{Bogolubsky:2007ud} 
 pure  Yang-Mills (no quarks included).

In this talk we will present recent results from a gauge-invariant study of 
the coupled gluon-ghost system of SDEs~\cite{Aguilar:2008xm}, yielding  
an infrared finite gluon propagator and a divergent (but non-enhanced) 
ghost propagator, in qualitative agreement with recent lattice data~\cite{Cucchieri:2007md,Bogolubsky:2007ud};
 this behavior has also been confirmed within the Gribov-Zwanziger formalism \cite{Dudal:2008sp}.
As the title suggests, we will eventually focus on the issue of the infrared finite QCD effective charge.
 

Obtaining an infrared finite result for the gluon self-energy from SDEs,
without violating the underlying local gauge symmetry, is far from trivial,
and hinges crucially on one's ability to devise
a  self-consistent truncation  scheme  that would  select a  tractable
and, at the same time, {\it physically meaningful} subset of these equations.  
Specifically, while in QED 
the Green's functions satisfy naive 
Ward Identities (WIs), in QCD they satisfy complicated 
Slavnov-Taylor identities (STIs), which 
involve various composite ghost operators. To see how this 
complicates the truncation procedure of the SDEs,
consider the STI of the gluon self-energy 
\be
q^{\mu} \Pi_{\mu\nu}(q) = 0 \,.
\label{fundtrans}
\ee
Eq.~(\ref{fundtrans}) is without a doubt the most fundamental 
statement at the level of Green's functions that one can obtain
from the BRST symmetry; it affirms the transversality of the 
gluon self-energy and is valid both  
perturbatively to all orders as well as non-perturbatively. 
The problem is that 
in the SDE governing $\Pi_{\mu\nu}(q)$ enter higher order 
Green's functions, 
namely the fully-dressed fundamental vertices of the theory, 
which satisfy complicated STIs.
Thus, whereas in QED the validity of Eq.~(\ref{fundtrans}) 
can be easily seen at  
the level of the SDE, simply because \mbox{
$q^{\mu}\Gamma_{\mu}(p,p+q) =e\left[ S^{-1}(p+q)-S^{-1}(p)\right]$}, in QCD 
proving Eq.~(\ref{fundtrans}) is very difficult, and requires the 
conspiracy of all full vertices appearing in the SDE.
Truncating the SDE naively usually amounts to leaving out some of these vertices, 
and, as a result, Eq.~(\ref{fundtrans}) is compromised.
Instead, the  gauge-invariant  truncation  scheme \cite{Binosi:2007pi},  
based  on  the  PT~\cite{Cornwall:1982zr, Cornwall:1989gv} and  its 
correspondence with the background field method (BFM)~\cite{Abbott:1980hw} 
maintains the validity of Eq.(\ref{fundtrans}) at every level of approximation.

\begin{figure}
\includegraphics[width=14cm]{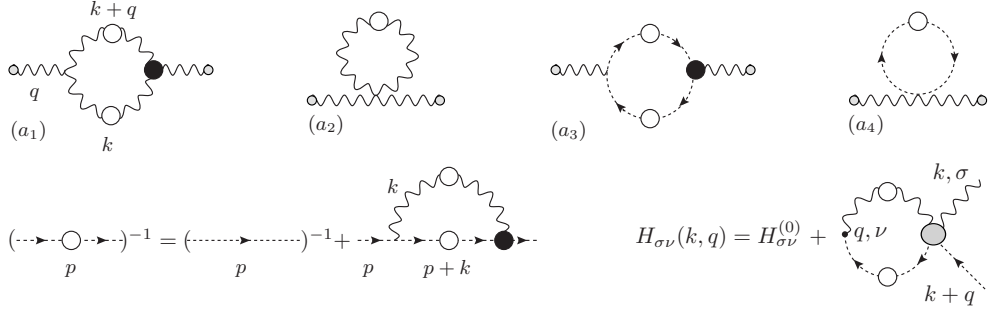}
\caption{The PT-BFM gluon-ghost system. The circles attached 
to the external gluons denote that, 
from the point of view of Feynman rules, they are treated as background fields.}
\label{SDeqs}
\end{figure}

The gluon propagator in the covariant gauges has the form
\mbox{ 
$i\Delta_{\mu\nu}(q)= \left[ {\rm P}_{\mu\nu}(q)\Delta(q^2) +\xi\frac{\D q_\mu
q_\nu}{\D q^4}\right]$},
where $\xi$ denotes the gauge-fixing parameter, 
\mbox{${\rm P}_{\mu\nu}(q)= g_{\mu\nu} - q_\mu q_\nu /q^2$}
is the usual transverse projector, and, finally, $\Delta^{-1}(q^2) = q^2 + i \Pi(q^2)$, 
with  $\Pi_{\mu\nu}(q)={\rm P}_{\mu\nu}(q) \,\Pi(q^2)$ the gluon self-energy. 
The full ghost propagator $D(p^2)$ and its self-energy $L(p^2)$ are related by
\mbox{$iD^{-1}(p^2)= p^2 - iL(p^2)$}. In the case of pure (quarkless) QCD,
the new SD series~\cite{Binosi:2007pi} for the gluon and ghost propagators reads (see also Fig.~\ref{SDeqs})
\bea
&& \Delta^{-1}(q^2){\rm P}_{\mu\nu}(q) = 
\frac{q^2 {\rm P}_{\mu\nu}(q) + i\,\sum_{i=1}^{4}(a_i)_{\mu\nu}}{[1+G(q^2)]^2} \,,
\nonumber\\
&& iD^{-1}(p^2) = p^2 +i \lambda  \int_k
\Gamma^{\mu}\Delta_{\mu\nu}(k)\gb^{\nu}(p,k) D(p+k) \,,
\nonumber\\
&& i\Lambda_{\mu \nu}(q) = \lambda 
\int_k H^{(0)}_{\mu\rho}
D(k+q)\Delta^{\rho\sigma}(k)\, H_{\sigma\nu}(k,q)\,,
\label{newSDb}
\eea
where \mbox{$\lambda=g^2 C_{\rm {A}}$}, with $C_{\rm {A}}$ the Casimir eigenvalue of the adjoint representation
[$C_{\rm {A}}=N$ for $SU(N)$], 
and \mbox{$\int_{k}\equiv\mu^{2\varepsilon}(2\pi)^{-d}\int\!d^d k$}, 
with $d=4-\epsilon$ the dimension of space-time. 
$\Gamma_{\mu}$ is the standard (asymmetric) gluon-ghost vertex at tree-level,
and $\gb^{\nu}$ the fully-dressed one.
$G(q^2)$ is the $g_{\mu\nu}$ component
of the auxiliary two-point function $\Lambda_{\mu \nu}(q)$, and the function $H_{\sigma\nu}$ is 
defined diagrammatically in Fig.~\ref{SDeqs}.
$H_{\sigma\nu}$ is related to the full gluon-ghost vertex by 
\mbox{$q^{\sigma} H_{\sigma\nu}(p,r,q) = -i{\gb}_{\nu}(p,r,q)$}; 
at tree-level, $H_{\sigma\nu}^{(0)} = ig_{\sigma\nu}$.
When evaluating  the diagrams $(a_i)$  we use the  BFM Feynman
rules~\cite{Abbott:1980hw}; the BFM fully dressed three-gluon and
gluon-ghost vertices are denoted by ${\g}_{\mu\alpha\beta}$ and ${\g}_{\mu}$.


Notice a point of paramount importance:
due to the Abelian all-order WIs that these two full 
vertices  satisfy (for all $\xi$), namely
\bea
q^{\mu}{\g}_{\mu\alpha\beta} = i\Delta^{-1}_{\alpha\beta}(k+q)
-i\Delta^{-1}_{\alpha\beta}(k)\,,\qquad
q^{\mu}{\g}_{\mu} = iD^{-1}(k+q) - iD^{-1}(k),
\label{PTWI}
\eea
one can demonstrate that 
$q^{\mu}[(a_1)+(a_2)]_{\mu\nu} =0$
and \mbox{$q^{\mu}[(a_3)+(a_4)]_{\mu\nu} =0$} \cite{Aguilar:2006gr}.
Thus, unlike other treatments in the literature, within this formalism 
the transversality of the gluon self-energy, i.e. Eq.~(\ref{fundtrans}), 
is preserved at every step, in absolute compliance with the BRST symmetry.

Next, following standard techniques,
we express ${\g}_{\mu\alpha\beta}$ and ${\g}_{\mu}$
as a function of the gluon and ghost self-energy, respectively, in such a way as to 
automatically satisfy the crucial WIs of  Eq.~(\ref{PTWI}); failure to satisfy these WIs 
would invariably compromise the transversality of the answer.  
The Ansatz we will use is 

\bea
{\g}_{\mu\alpha\beta}= \Gamma_{\mu\alpha\beta} + i\frac{q_{\mu}}{q^2}
\left[\Pi_{\alpha\beta}(k+q)-\Pi_{\alpha\beta}(k)\right]\nonumber \,,  \qquad
{\g}_{\mu}= \widetilde{\Gamma}_{\mu} -i\frac{q_{\mu}}{q^2}
\left[L(k+q)-L(k)\right]\,;
\label{gluonv}
\eea
its essential feature, other than satisfying the aforementioned WIs, 
is the presence of massless, longitudinally coupled   
pole terms, which are instrumental for obtaining 
$\Delta^{-1}(0)\neq 0$~\cite{Jackiw:1973tr}. These poles are not kinematic but dynamical, 
corresponding to a  composite (bound-state) Goldstone excitation,  enforcing the local gauge invariance.
For the conventional ghost-gluon vertex ${\gb}_{\nu}$,
appearing in the second SDE of (\ref{newSDb}) we will use its
tree-level expression, {\it i.e.}, ${\gb}_{\nu}\to \Gamma_{\nu} = -p_{\nu}$;
this is perfectly legitimate, since in this formalism 
the two ghost vertices,  ${\gb}_{\nu}$ and ${\g}_{\mu}$, are different.
Finally, for $H_{\sigma\nu}$ we use its tree-level value, $H_{\sigma\nu}^{(0)}$.

\begin{figure}[ht]
\hspace{-1.5cm}
\includegraphics[scale=2.0]{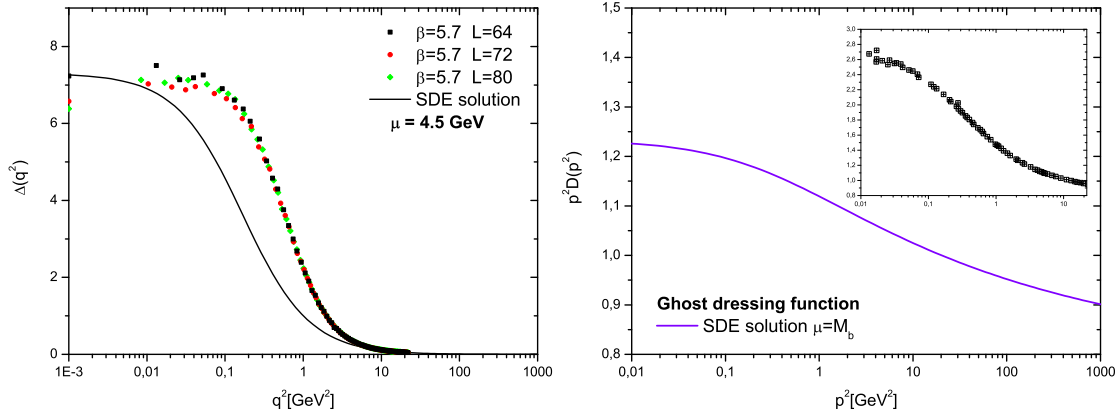}
\vspace{-2.0cm}
\caption{Left Panel:The numerical solution for the gluon propagator from the SDE
(black continuous line) compared to the 
lattice data of Ref.\cite{Bogolubsky:2007ud}. Right panel:The ghost dressing function $p^2D(p^2)$ 
obtained from the SDE. In the insert we show  
the lattice data for the same quantity; notice that there is no ``enhancement''\cite{Boucaud:2008ky}.}
\label{figb}
\end{figure}

In Fig.\ref{figb}, we show the numerical result for  $\Delta(q^2)$  
renormalized at $\mu=M_b=4.5\,\mbox{GeV}$, and the comparison 
with the corresponding lattice data of Ref.\cite{Bogolubsky:2007ud}.
In the right panel of Fig.\ref{figb}, we present the dressing function 
for the ghost propagator, renormalized at the same point.
\vspace{-1.5cm}


Let us now turn to the QCD effective charge. There are two main issues:
(i) how to define it consistently at the level of perturbation theory:
specifically, which graphs determine the running, and 
(ii) how to extend the (whatever) definition one reaches in (i) 
into the non-perturbative regime.

Point (i) has been addressed exhaustively in the literature~\cite{Watson:1996fg}: the upshot  is that in the 
context of the PT one may replicate to all-orders in perturbation theory the prototype QED construction 
of an effective charge. To fix the ideas,  the PT one-loop gluon self-energy 
reads
\be 
\widehat\Delta^{-1}(q^2)= q^2\left[1+ b g^2\ln\left(\frac{q^2}{\mu^2}\right)\right]\,,
\label{rightRG}
\ee
where  $b = 11 C_A/48\pi^2$  is the first coefficient of the QCD $\beta$-function. 
Due to the Abelian WIs satisfied by the PT effective Green's functions, the 
 new propagator-like quantity $\widehat\Delta^{-1}(q^2)$ absorbs all  
the RG-logs, exactly as happens in QED with the photon self-energy.
Equivalently, since $Z_{g}$ and ${\widehat Z}_{A}$, the renormalization constants 
of the gauge-coupling and the effective self-energy, respectively, 
satisfy the QED relation ${Z}_{g} = {\widehat Z}^{-1/2}_{A}$,  
the product 
${\widehat d}(q^2) = g^2 \widehat\Delta(q^2)$ forms a RG-invariant 
($\mu$-independent) quantity~\cite{Cornwall:1982zr};
for large momenta $q^2$,
\be
{\widehat d}(q^2) = \frac{\overline{g}^2(q^2)}{q^2}\,,
\label{ddef1}
\ee
where $\overline{g}^2(q^2)$ is the RG-invariant effective charge of QCD,
\be
\overline{g}^2(q^2) = \frac{g^2}{1+  b g^2\ln\left(q^2/\mu^2\right)}
= \frac{1}{b\ln\left(q^2/\Lambda^2\right)}\,.
\label{effch}
\ee
%
\begin{figure}[ht]
\vspace{-1.5cm}
\hspace{-2.0cm}
\includegraphics[scale=2.0]{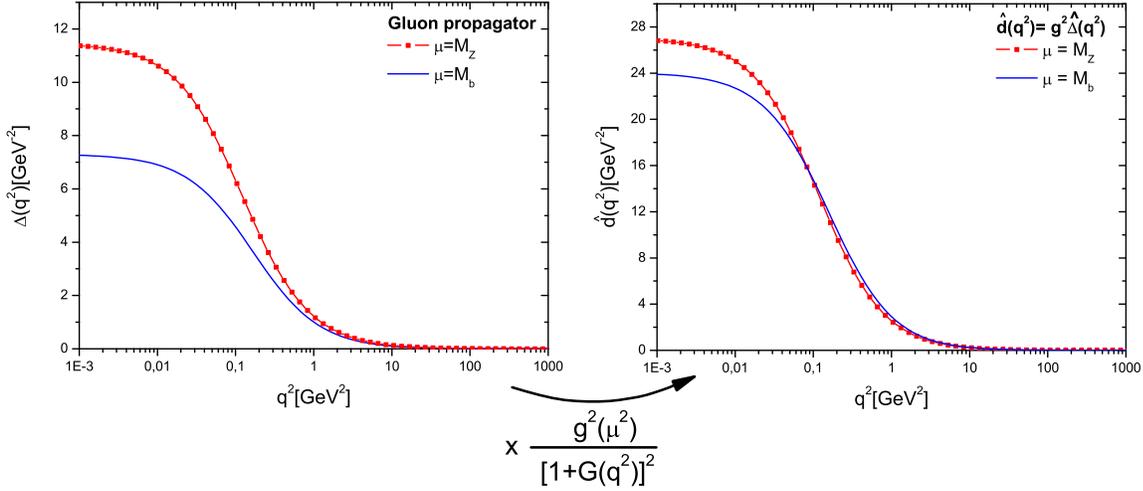}
\vspace{-1.2cm}
\caption{Left panel: The solution of SDE renormalized at \mbox{$\mu=M_b=4.5\,\mbox{GeV}$}
(continuous blue curve) and  $\mu=M_Z=91\,\mbox{GeV}$ (red line+square curve). Right Panel: The
 corresponding PT-BFM $\widehat{\Delta}(q^2)$ obtained as the 
convolution of $\Delta(q^2)$  and the function $g^2(\mu^2)/[1+G(q^2)]^2$. }
\label{rgi_fig}
\end{figure}
%
Let us now come to point (ii): assuming that one has non-perturbative 
information about the infrared behavior of the conventional gluon propagator $\Delta(q^2)$, 
how should one extract an effective charge, which, perturbatively, 
will go over to Eq.~(\ref{effch})? To accomplish this, one must use an additional 
field-theoretic ingredient:
the conventional $\Delta(q^2)$ and the PT-BFM $\widehat{\Delta}(q^2)$
are related by the formal all-order relation~\cite{Binosi:2002ez} 
\be
\Delta(q^2) = 
\left[1+G(q^2)\right]^2 \widehat{\Delta}(q^2) \,.
\label{bqi}
\ee
Note that the  $G(q^2)$  already appears in  Eq.~(\ref{newSDb}) and Fig.\ref{SDeqs}.
With our approximations its SDE reads
\be
G(q^2) = - \frac{\lambda}{3}\int_k\,\left[
2+\frac{(k\cdot q)^2}{k^2q^2}\right]\Delta(k)D(k+q) \,.
\label{gg}
\ee

First of all, it is easy to verify that at lowest order 
the $G(q^2)$ obtained from  Eq.~(\ref{gg})  restores the $\beta$ function coefficient  
in front of ultraviolet logarithm. In that limit 
\mbox{$1+G(q^2) = 1 +\frac{9}{4}
\frac{C_{\rm {A}}g^2}{48\pi^2}\ln(q^2/\mu^2) $} and \mbox{$\Delta^{-1}(q^2)= q^2 \left[1+\frac{13}{2}
\frac{C_{\rm {A}}g^2}{48\pi^2}\ln(q^2/\mu^2)\right]$}. Then using  
Eq.~(\ref{bqi}) we recover the $\widehat{\Delta}^{-1}(q^2)$ of Eq.~(\ref{rightRG}), as we should. 
Then, non-perturbatively, one substitutes into  Eq.~(\ref{bqi}) the $G(q^2)$ and $\Delta(q^2)$ 
obtained from solving the system in  Eq.~(\ref{newSDb}), to obtain $\widehat{\Delta}(q^2)$.  
This latter quantity is the non-perturbative generalization of Eq.~(\ref{rightRG}); 
for the same reasons explained above, when multiplied by $g^2$ it should form an RG-invariant quantity, e.g. the non-perturbative 
generalization of ${\widehat d}(q^2)$. In Fig.\ref{rgi_fig} we present the combined result of the above steps: 
${\widehat d}(q^2)$ is obtained from two different sets of solutions of the system  Eq.~(\ref{newSDb}), one  
renormalized at $\mu=M_b=4.5\,\mbox{GeV}$ and one at $\mu=M_Z=91\,\mbox{GeV}$. Ideally
the two curves 
of ${\widehat d}(q^2)$ should be identical; even though this does not happen, due to the approximations employed 
when  solving the system of Eq.~(\ref{newSDb}), the two curves are fairly close, 
indicating that ${\widehat d}(q^2)$ is to a very good approximation an RG-invariant quantity, as it should.

We are now in the position to define the non-perturbative 
QCD effective charge from the RG-invariant quantity ${\widehat d}(q^2)$. Of course, given that 
${\widehat d}(q^2)$ reaches a finite value in the deep infrared, 
it would be completely absurd to define the effective charge 
by  forcing out a factor of $1/q^2$; such a procedure would furnish a completely unphysical strong QCD coupling, 
namely one that would vanish in the deep infrared(!) where QCD is supposed to be strongly coupled.
\vspace{-1.1cm}
\begin{figure}
\begin{minipage}[c]{.45\textwidth}
\hspace{-1.0cm}
\includegraphics[scale=0.8]{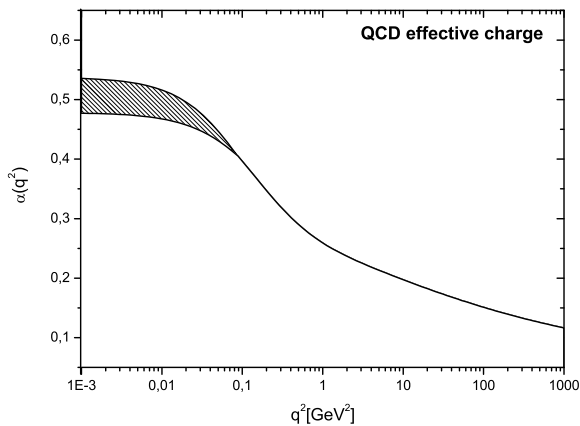}
\vspace{-1.4cm}
\begin{center}
\caption{The QCD effective charge, \mbox{$\alpha(q^2)=\overline{g}^2(q^2)/4\pi$},
extracted from Fig.3 by factoring out a gluon mass of 
\mbox{m(0)=500\, \mbox{MeV}}.}
\label{figc}
\end{center}
\end{minipage}
\hspace{\fill}
\begin{minipage}[l]{.5\textwidth}
{The correct thing to do is to factor out a ``massive'' propagator, i.e. write 
\be
\widehat{d}(q^2) = \frac{\overline{g}^2(q^2)}{q^2 + m^2(q^2)} \,.
\label{ddef}
\ee

Of course, as we have emphasized,  $m^2(q^2)$ itself is running, which must also 
be taken into account in a more sophisticated treatment.
For the purposes of this talk, however, we assume that 
$m^2(q^2)$ is constant, $m^2(q^2)=m^2(0)$, and use for $m(0)$ the value of $500\, \mbox{MeV}$ 
favored by phenomenology~\cite{Aguilar:2001zy}. 
The $\alpha(q^2)$ obtained is shown in Fig.(\ref{figc}); as announced, 
at low energies freezes to a finite value, indicating the appearance of an infrared fixed point of QCD.
 }
\end{minipage}
\end{figure}
%

\acknowledgments
 We thank Prof. J.~M.~Cornwall for numerous illuminating discussions.
Work supported by the Spanish MEC grants FPA 2005-01678 and the Fundaci\'on General of the UV.

\end{document}